\newcounter{myctr}
\def\myitem{\refstepcounter{myctr}\bibfont\noindent\ifnum\themyctr>9\else\phantom{0}\fi\hangindent17pt\themyctr.\enskip}

\documentclass{ws-ijqi}
\usepackage{amssymb}
\usepackage{graphicx}
\usepackage{comment}
\usepackage{mathtools}
\usepackage{bbold}
\usepackage[titletoc]{appendix}
\usepackage[dvipsnames]{xcolor}
\usepackage{courier}
\usepackage{hyperref}
\usepackage{xcolor}
\usepackage{xpatch}
\usepackage[normalem]{ulem}
\usepackage{hyperref}
\usepackage[super,sort,compress]{cite}
\usepackage{graphics}
\usepackage{bm}
\usepackage{bbm}
\usepackage{amsbsy}

\usepackage{babel}
\definecolor{airforceblue}{rgb}{0.36, 0.54, 0.66}
\definecolor{darkolivegreen}{rgb}{0.33, 0.42, 0.18}
\definecolor{cerulean}{rgb}{0.0, 0.48, 0.65}

\hypersetup{
    colorlinks=true,
    linkcolor=cerulean,
    filecolor=magenta,
    citecolor=darkolivegreen,
    urlcolor=airforceblue,
    pdftitle={Overleaf Example},
    pdfpagemode=FullScreen,}

\usepackage{float}

\DeclareMathOperator{\Tr}{Tr}

\newcommand{\mbraket}[1]{\langle #1 \rangle}

\newcommand{\HC}{\mathcal{H}}

\newcommand{\LC}{\mathcal{L}}

\newcommand{\NC}{\mathcal{N}}

\newcommand{\QC}{\mathcal{Q}}

\newcommand{\SC}{\mathcal{S}}

\begin{document}

\catchline{}{}{}{}{}

\title{Revisited aspects of the local set in CHSH Bell scenario}

\author{Nicol\'as Gigena}
\address{Faculty of Physics, University of Warsaw, Pasteura 5, 02-093 Warsaw, Poland}
\author{Giovanni Scala}
\address{International Centre for Theory of Quantum Technologies, University
of Gdansk,  Jana Ba\.zy\'nskiego 1A, 80-309 Gdansk, Poland}
\author{Antonio Mandarino}
\address{International Centre for Theory of Quantum Technologies, University
of Gdansk,  Jana Ba\.zy\'nskiego 1A, 80-309 Gdansk, Poland}

\maketitle

\begin{abstract}
\noindent The Bell inequalities stand at the cornerstone of the developments of quantum theory on both the foundational and applied side. 
The discussion started as a way to test whether the quantum description of reality is complete or not, 
but it developed in such a way that a new research area stemmed from it, namely quantum information. Far from being and exhausted topic, 
in the present paper we present a constructive and geometrically intuitive description of the local polytope and its facets 
in a bipartite Bell scenario with two dichotomic measurements per party.
\end{abstract}
\keywords{Bell inequalities, CHSH }

\section{Introduction}

The question of whether a broader theory that recovers the statistics of quantum measurements 
via average on local variables bemused some of the founding fathers of quantum mechanics. 
Anyway, it was the work of John Bell\cite{Bell1964,Bell1964a} and the subsequent formalization in form of inequalities 
involving the expectation values of operators acting on two separate systems that marked the turning point from mere speculation to an impossibility-proof experimentally testable. This was achieved by the \emph{CHSH inequality} a work of Clauser, Horne, Shimony, and Holt\cite{CHSH} and later modified and extended by the \emph{CH inequality}, a work of
Clauser and Horne\cite{CH74} (a recent algebraic variant of CHSH is in the work of Bavarian and Shor\cite{Bavarian2015}). Both inequalities refer to a scenario in which two observers share a pair of objects on which they perform two different and \textit{arbitrary} spatially separate measurements each on the object of the pair and collect a series of dichotomic results. In this manuscript, by \emph{classical} we mean a local realistic theory: measurement outcomes already exist before the measurement and are independent of it and from the presence of an observer (realism); the locality assumption implies that the outcomes of one part do not depend on the settings of the other one.

This assumption is usually called \emph{Bell locality} to avoid confusion with the \emph{non-signaling} assumption. A violation of a Bell inequality might infer that \emph{Bell locality} is violated, however \emph{non-signaling} assumption is always satisfied, causing the theory to be local, but not Bell local. Therefore, we will say that quantum mechanics is not Bell–local, but it is a local theory because does not violate the no-signaling condition.
Overall, the difference between the two formulations, CHSH, and CH, is given by the objects used to define the 
 inequalities satisfied by any local realistic theory and violated by \emph{some} quantum predictions. The Bell-CHSH inequality makes use of the average of two-point correlation functions evaluated on the chosen experimental settings for the first experimenter and the second one. Then, the average is taken over many experimental runs for the local hidden variable model or 
over the required quantum state. In contrast, the CH inequality makes use of the local event probabilities and joint probabilities for events detected at each observer side. In the following, we will be focused on CHSH inequality.

\section{Bipartite Bell scenarios, Correlation sets and Bell inequalities \label{sec:sets}}

A bipartite Bell scenario 
consists of two physical systems, spatially separated, which can be locally accessed by two observers, Alice and Bob. Alice is allowed to perform, at a time, one out of $m_A$ different measurements, each yielding $d_A$ possible outcomes. Analogously, Bob can perform one out of $m_B$ measurements, having each a total of $d_B$ possible outcomes. This four numbers completely characterize the scenario, so we can label scenarios by four number tuples $(m_A, m_B, d_A, d_B)$, or just $(m, d)$ when the numbers of settings and outcomes are the same for both parties.

In the following, we will label Alice's (Bob's) measurement settings with $x (y)$, and use $a (b)$ to denote their possible outcomes. Both Alice and Bob can agree in performing a sequence of measurements on their respective system, choosing different measurement settings each time so that after many runs of their protocol they can come up with an estimate of the probabilities for the outcomes given a particular choice of setting. There are a total of $m_A d_A m_B d_B$ such probabilities, that we will denote by $p(ab|xy)$, in accordance with the labels we have chosen. It is convenient to arrange these probabilities in a tuple $\bm p$, that we will think of as a point in $\mathbb{R}^{m_A d_A m_B d_B}$ and to which we will refer as a {\it behavior}. Note that not every point in the real vector space corresponds to one of these probability points, because for the latter the components are constrained to belong to the $[0, 1]$ interval, and to satisfy the normalization constraint $\sum_{ab} p(ab|xy)=1\;\forall x, y$
. Moreover, because of the constraints, the dimensionality of the set of behaviors is necessarily lower than $m_A d_A m_B d_B$; we will come back to this issue in what follows.

\subsection{The no-signaling set}

The probability constraints given above are not the only ones to impose on the behaviors. Since these probabilities result from measurements performed on physical systems we would expect other, physically motivated constraints, to also be in order. The most obvious of these is the no-signaling constraint: communications or signals faster than light are forbidden by the axioms of special relativity under which our physical systems are defined. Because we assume the measurement of Alice and Bob to be events spatially separated, it follows that the marginal probabilities estimated by Alice for her measurement outcomes should not be affected by Bob's choice of measurement setting, and vice-versa. These no-signaling conditions are imposed on the behaviors by requiring the following relations to be satisfied \cite{Peres2004,Barrett,Cope2019}
\begin{equation}
\begin{split}
    \sum_{b} p(ab|xy) &= \sum_{b} p(ab|xy') \qquad\forall y, y', \\
    \sum_{a} p(ab|xy) &= \sum_{a} p(ab|x'y) \qquad\forall x, x'.
\end{split}
\end{equation}
It is not hard to see that the set $\mathcal{NS}$ of behaviors satisfying these constraints as well as the previously introduced positivity constraints is convex -- it is enough to prove that if any two different behaviors satisfy the constraints, then any convex combination of them will also satisfy them. Moreover, because the constraints are linear it follows that this set is the intersection of a finite set of half-spaces, i.e., a convex polytope \cite{Pironio2014,Brunner2014,Ziegler1995}. We can alternatively think of this polytope as the convex hull of a set of extremal points, which is commonly known as {\it vertex representation}\cite{Green1958}. These extremal points, or vertices, are unique solutions to a subset of constraint equations, and the problem of finding them is known as the {\it vertex enumeration problem}\cite{Avis1992, Weisstein2002, Khachiyan2008}. As stated before, the constraints we impose will reduce the dimensionality of the polytope, and the no-signaling constraints are no exception. One way to realize this dimension reduction is to consider the set of probabilities $\{p(a|x), p(b|y), p(ab|xy)\}$, where $a (b) =0, 1, ..., d_A -2 (d_B-2)$. After this parametrization, which takes into account the no-signaling and normalization constraints, the no-signaling behaviors are only constrained by the positivity condition $p(ab|xy)\geq 0$, and restricted to a real subspace of dimension\cite{Pironio2014}
\begin{equation}
    D=(d_A-1)m_A+(d_b-1)m_B+(d_A-1)(d_B-1)m_Am_B. \label{eq:efdim}
\end{equation}

\subsection{The local set}
There is another physically motivated constraint that we impose on our behaviors: The locality constraint. In words, when we call a behavior local we are saying that all correlations between Alice's and Bob's output come from a shared resource. In a more formal way, this means that for any local behavior we can find a variable --or a finite set of them-- $\lambda$, distributed according a probability distribution $\rho(\lambda)$ allowing the joint probabilities $p(ab|xy)$ to be decomposed as
\begin{equation}
    p(ab|xy)=\int {d\lambda}\, \rho(\lambda) p(a|x, \lambda)p(b|y\lambda) \label{eq:loc}   
\end{equation} 
A variable such as $\lambda$ is commonly referred to as a {\it hidden} or {\it latent} variable, and behaviors satisfying condition \eqref{eq:loc} are said to admit a local hidden variable model.  It can be proved, as was done by Fine in 1982\cite{Fine1982}, that any local behavior can be expressed as a convex combination of deterministic local behaviors, i.e., behaviors for which the marginal conditional probabilities $p(a|x, \lambda), p(b|y, \lambda)$ take values in the set $\{0, 1\}$. The intuition behind this result comes from realizing that any randomness present in these marginal distributions can always be absorbed by the hidden variable $\lambda$. Because of this, it turns out that the set $\LC$ of local behaviors is the convex hull of the finite set of deterministic behaviors, i.e., a polytope. Moreover, since the behaviors in the local set satisfy by construction the constraints defining the $\NC\SC$, it follows that $\LC\subset\NC\SC$, and it has the same dimension \cite{Pironio2005,Barrett1}.

It can be shown that the local set is strictly smaller than the no-signaling set, that is, there exist behaviors in $\NC\SC$ that do not belong to $\LC$ \cite{PR_orig}. Behaviors $\bm p$ satisfying $\bm p\in\NC\SC$ and $\bm p\notin \LC$ are usually referred to as {\it nonlocal}. From a geometrical point of view, what this means is that while some faces of $\LC$ are also faces of $\NC\SC$ --those defined by the positivity constraint-- other faces will be the boundary between the local behaviors and those behaviors that satisfy the no-signaling constraint, but violate the locality constraint.

\subsection{The quantum set} 

There is yet another physically motivated constraint that we could impose on the behaviors we are considering: we can demand from the joint probabilities $p(ab|xy)$ to admit a quantum realization. Formally, this means that we ask for the existence of a composite quantum state $\rho$ and local measurement operators $\{M_x^a, M_y^b\}$ such that the joint probabilities can be written as
\begin{equation}\label{qprob}
    p(ab|xy)=\Tr \rho M_x^a\otimes M_y^b.
\end{equation}
Here the operators $M_{x}^{a}$ and $M_{x}^{a}$ are assumed to act on a local Hilbert space $\HC_A$ and $\HC_B$ respectively, whereas the state $\rho$ is an operator over the joint Hilbert space $\HC_A\otimes\HC_B$. Thanks to Naimark's theorem we can consider without loss of generality that the measurements are projective, i.e., $M_{x}^{a}M_{x}^{a'}=\delta_{a a'}M_{x}^{a}$ and $M_{y}^{b}M_{y}^{b'}=\delta_{bb'}M_{y}^{b}$, and also we can take the state $\rho$ to be pure.

The set $\QC$ of probabilities admitting a quantum realization is, like $\NC\SC$ and $\LC$, a convex set. But unlike these, it is not a polytope, which makes of the description of its boundary a hard task. Moreover, it is easy to see that any behavior within the local set admits a quantum realization \cite{Brunner2014}, but as we will see in what follows there are quantum behaviors that violate the locality condition. Thus we have that $\LC\subset\QC$, i.e., the local set is a strict subset of the quantum set. On the other hand, we can easily show that quantum behaviors satisfy the no-signaling constraint  \footnote{This follows directly from the completeness relation $\sum_{a (b)} M_{x (y)}^{a (b)}=I_{A (B)}$, with $I_{A (B)}$ the identity operator on space $\HC_{A (B)}$, that the operators $M_{x (y)}^{a (b)}$ must satisfy in order to describe suitable quantum measurements.}, but there are no-signaling behaviors that do not admit a quantum realization, implying that $\QC$ is a strict subset of $\NC\SC$.

\subsection{Bell inequalities}

It follows from the discussion above that there is a strict inclusion relation between the correlation sets, i.e., 
\begin{equation}
    \LC\subset\QC\subset\NC\SC
\end{equation}
The first inclusion is of particular interest, since from it follows that there are quantum probability points not admitting a local-realistic explanation. In other words, it shows that there is indeterminacy in nature \cite{scarani2019bell,Wood2015,chiribella2010probabilistic}. The existence of such behaviors, a phenomenon commonly known as {\it Bell nonlocality}, is not only of paramount importance from a foundational point of view: It has also practical consequences since it plays a key role in quantum information processing tasks like device-independent quantum key distribution protocols and randomness certification \cite{acin2007device, masanes2011secure}.

The natural question is therefore, given a quantum behavior $\bm q$, how can we decide whether it is a nonlocal behavior? To answer this question we can use the Hahn-Banach theorem, also known as hyperplane separation theorem \cite{kolmogorov1957elements}. Because the local set $\LC$ is convex, the theorem states that for any nonlocal behavior $\bm q$ in $\QC$ there exists a hyperplane that separates it from $\LC$. As a result, there exists a linear functional $\bm\beta:\mathbb{R}^D\mapsto \mathbb{R}$ and a real number $\beta_c$ such that $\bm \beta\cdot\bm p\leq \beta_c\,\forall \bm p\in\LC$, and $\bm\beta\cdot\bm q>\beta_c$. The first of these inequalities is usually referred to as {\it Bell inequality}. Because the local set is the intersection of a finite set of halfspaces and it is also a strict subset of the no-signaling set, it follows that there is a finite set of $D-1$ dimensional faces (or {\it facets}) of the local polytope that serve as separating hyperplanes for {\it all} nonlocal behaviors. The Bell inequalities associated with these facets, known as \emph{tight} Bell inequalities, are obviously of special importance since their knowledge gives a simple solution of the membership problem discussed above: a behavior $\bm p$ is local if it satisfies all tight Bell inequalities. 

While it is true that the knowledge of all non-trivial facets of the local polytope gives a simple solution to the membership problems, in practice the difficulty is not removed but just relocated, since we normally deal with the local set in its vertex representation. The reason for this is that these vertices correspond to local deterministic strategies, which are easy to characterize. Therefore, in order to be able to check the locality of a given behavior we first need to find the non-trivial facets of the local polytope. The problem of obtaining the half-space representation of a polytope from its vertex representation is known as {\it facet enumeration problem} \cite{Chaves2016}, which is in general very hard. In what follows we will explore this problem for in the simplest bipartite Bell scenario, consisting of $2$ dichotomic measurements per party.


\section{The local polytope in the (2, 2) scenario}

The $(2, 2)$ scenario is the simplest Bell scenario in which the nonlocal nature of quantum correlations can manifest. As indicated by the label $(2, 2)$, Alice and Bob can perform only two dichotmic measurements on their systems. Let Alice and Bob label their two possible measurement settings with $x, y\in\{0, 1\}$ respectively, and the outcomes of these measurements with $a, b\in I=\{-1, 1\}$. A behavior in this scenario is therefore determined by $16$ probabilities $p(ab|xy)$. 

Local behaviors, as stated in the previous section, can always be written as convex combinations of local deterministic points. Combining the locality condition in Eq.~\eqref{eq:loc} with the eterminism constraint, i.e., $p(a|x), p(b|y)\in\{0, 1\}$ we get that deterministic behaviors are those satisfying the condition $p(ab|xy)=p(a|x)p(b|y)\,\forall a, b, x, y$. There are $2^4=16$ of such probability points in this scenario, and therefore the local set $\LC$ is a $16$-vertex polytope in an $8$-dimensional real space, as follows from Eq.~\eqref{eq:efdim}. Since the measurements in this scenario are dichotomic, instead of dealing with probabilities we can describe the vertices of $\LC$ using correlators, linear functions of the probabilities defined as
\begin{eqnarray}
        \mbraket{A_xB_y}&=&\sum_{a,b\in I} ab \,p(ab|xy),\label{2points}\\
        \mbraket{A_x}=\sum_{a\in I} a\, p(a|x),&\quad &\mbraket{B_y}=\sum_{b\in I} b \,p(b|y).\label{1point}\quad
\end{eqnarray}
where $\mbraket{A_xB_y}$, $\mbraket{A_x}$ and $\mbraket{B_y}$ take values in $[-1, 1]$. In terms of these correlators we can express any local deterministic behavior as a pair $\bm v=(\bm m, \bm c)\in\mathbb{R}^4\oplus\mathbb{R}^4$, with components given by
\begin{equation}
\begin{split}
    \bm m &=(\mbraket{A_0}, \mbraket{A_1}, \mbraket{B_0}, \mbraket{B_1}), \\ 
    \bm c &=(\mbraket{A_0}\mbraket{B_0}, \mbraket{A_0}\mbraket{B_1}, \mbraket{A_1}\mbraket{B_0}, \mbraket{A_1}\mbraket{B_1}) \label{eq:vector}
\end{split}
\end{equation}
with each component taking values in $\{-1, 1\}$. It is clear from these definitions that the $16$ vertices come in 8 pairs $\bm v_i^\pm=(\pm\bm m_i, \bm c_i)$, $i=1, ..., 8$. We are now interested in finding the non-trivial facets of the polytope determined by these $16$ local deterministic behaviors. In the general case these can be found numerically, implementing an algorithm that solves the facet enumeration problem, but since this is the simplest scenario we propose instead a more heuristic exploration of the problem, intended to build some intuition about the geometric features of $\LC$.

We begin by noting that the correlation part of the local vertices $\bm v_i^{\pm}$ is invariant under a swap of both Alices's and Bob's measurement outputs, i.e., $a\Rightarrow a+1 \mod 2$, $a\Rightarrow a+1 \mod 2$. We can ask then whether there are Bell inequalities that share this symmetry, since in that case they should be associated to functionals $\bm\beta:\mathbb{R}^8\mapsto \mathbb{R}$ of the form $\bm\beta=(\bm 0,  \bm\beta_c)$, because $\bm\beta\cdot\bm v_i^+=\bm\beta\cdot\bm v_i^-$ must hold for at least one of the $8$ vertex pairs. It follows therefore that any nonlocal behavior that can potentially be detected by an inequality as such should have a correlation component that lies outside the four-dimensional polytope 
\begin{equation}
    \LC_{c}=\{\bm c\in\mathbb{R}^4 \,|\, \bm c=\sum_i\ \lambda_i\bm c_i, \,\lambda_i\geq 0 \wedge \sum_i\lambda_i=1 \}\subset\mathbb{R}^4. \label{eq:corp}
\end{equation}
The $8$ vertices defining $\LC_c$ in Eq.~\eqref{eq:corp}, which form a $4$-dimensional octahedron, can be further classified into 2 different classes determined by the values of $\mbraket{A_0}$ and $\mbraket{B_0}$ --note that the choice is arbitrary, any other pair is equally good. Indeed, note that the marginal correlators for deterministic behaviors can be rewritten as
\begin{equation}\label{phase}
\begin{split}
    \mbraket{A_x}&=\mbraket{A_0}\,(-1)^{x r_A}, \\
    \mbraket{B_y}&=\mbraket{B_0}\,(-1)^{y r_B},
\end{split}
\end{equation}
with $r_A, r_B\in\{0, 1\}$. Using this notation we can write a given vertex of $\LC_c$ as
\begin{equation}
\begin{split}
    \bm c_{r_Ar_B}^s=s (1, (-1)^{r_B}, (-1)^{r_A}, (-1)^{r_A+r_B}), \label{eq:class}
\end{split}
\end{equation}
where $s\in\{-1, 1\}$ indicates whether $\mbraket{A_0} \mbraket{B_0}$ is positive or negative. Thus after fixing $s$ we are left with only $4$ vertices of the form
\begin{equation}
\label{joint_ind}
    \bm c_{r_A r_B}=(1, (-1)^{r_B}, (-1)^{r_A}, (-1)^{r_A+r_B}),
\end{equation}
or more explicitly,
\begin{align}
    \bm c_{00}&=(1, 1, 1, 1),\,& \bm c_{01}&=(1, -1, 1, -1)\nonumber\\
    \bm c_{11}&=(1, -1, -1, 1),\,& \bm c_{10}&=(1, 1, -1, -1), \label{tetrav}
\end{align}
which are easily seen to correspond to the vertices of a tetrahedron in the three dimensional subspace $\{\bm c\in\mathbb{R}^4 \,|\, \bm c=(1, x, y, z),\; x, y, z\in\mathbb{R}\}$.

\begin{figure}
	\centering
\includegraphics[width=.50\textwidth]{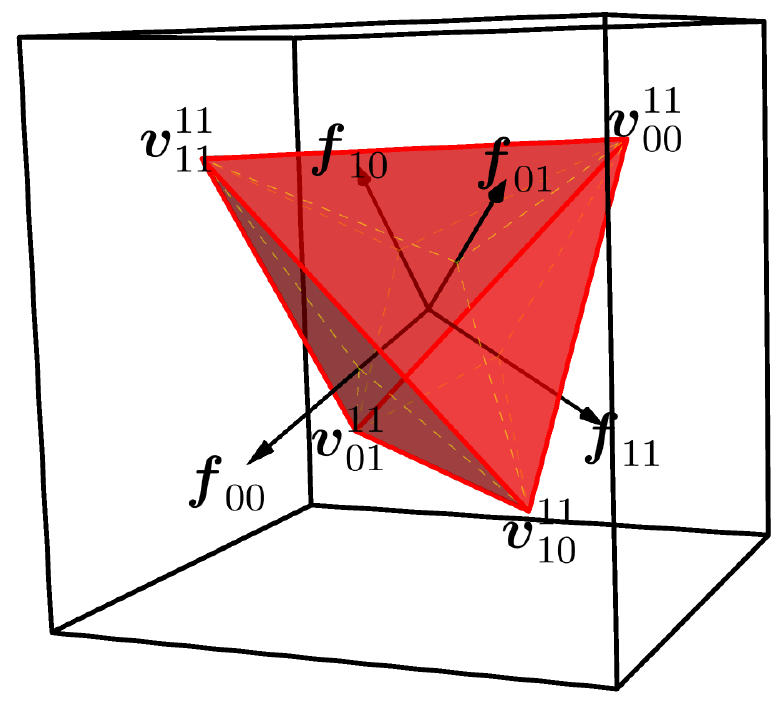}    
	\caption{The tetrahedron defines all the points $\LC$ in the subpolytope 
 at fixed $\mbraket{A_0},\mbraket{B_0}=1$. The vectors $\bm f_{ij}$ are normal to the 4 facets of the correlations and are pointing in the opposite direction of $v_{ij}$(the lighting yellow line helps to visualize it). The animated version of the picture is published in this  \href{https://github.com/giovanniscala/Bell_inequalities.git}{github folder}. A graphical representation of the relation between $\LC\subset\mathcal{Q}\subset\mathcal{NS}$ is in Ref.\cite{Pironio2014}.
	}
	\label{polytope}
\end{figure}

As stated above, if a nonlocal behavior is to be detected by our symmetric Bell inequalities then it should lie outside $\LC_c$, which means it must lie outside the two tetrahedrons we just described. This requires us again to solve a membership problem, but now our polytope is just a three-dimensional tetrahedron, and finding a solution is straightforward. Indeed, it is clear from the plot in Fig.~\ref{polytope} that the normal to a given face of the tetrahedron, in the outwards direction, is defined by the negative of the vertex opposed to it, so for the $8$ faces we find
\begin{equation}
    \begin{split}
        \bm f_{00}^s&=s (1, -1,- 1,-1), \\
        \bm f_{01}^s&=s (1, -1, -1, 1), \\
        \bm f_{10}^s&=s (1, -1, 1, 1), \\
        \bm f_{11}^s&=s (1, 1, 1, -1). \label{eq:cfunc}
    \end{split}
\end{equation}
As can be easily checked, $\bm f_{ij}^s\cdot\bm c^s_{r_Ar_B}=2$ whenever $(r_A, r_B)\neq (i, j)$. Hence we have the solution to our reduced membership problem: a correlation point $\bm c$ belongs to $\LC$ if it satisfies the $8$ linear inequalities $\bm f_{ij}^s\cdot\bm c\leq 2$. Furthermore, it is clear from this solution that these faces are nontrivial since the $8$ correlation points in Eq.~\eqref{eq:cfunc} correspond to no-signaling distributions --the positivity condition is satisfied--, and yet do not belong to $\LC_c$. It can be proved that this points, known in the literature as Popescu-Rorlich boxes, actually correspond to the $8$ nonlocal vertices of the $\NC\SC$ polytope \cite{Barret2005}. It follows from our construction that these $8$ nonlocal vertices toghether with the $8$ local deterministic vertices of $\LC_c$ define a four-dimensional hypercube, or {\it tesseract} \footnote{Recall that a four-dimensional hypercube is the convex hull of the vertices $(\pm 1, \pm 1, \pm 1, \pm 1)$.}.

We can now go back from the correlation subspace and consider the entire local polytope. Here the Bell functionals that we have found will have the form $\bm\beta_{ij}^s=(\bm 0, \bm f_{ij}^s)$, each associated to a nonlocal vertex of $\NC\SC$. The faces of $\LC$ associated to these functionals are easily seen to have dimension $7$, i.e., they are facets of $\LC$, since they are three-dimensional in the correlation subspace and do not have support on the four-dimensional marginal subspace. As a result, the functionals $\{\bm\beta_{ij}^s\}$ give rise to $8$ tight Bell inequalities. These facets are convex combinations of local vertices of which we already know the correlation components, which in turn toghether with Eq.~\eqref{eq:vector} determine the two possible marginal components. Hence, associated to the Bell functional $\bm\beta_{i, j}^s$ we have $6$ vertices of $\LC$ which are given by
\begin{equation}
    \bm v_{r_A r_B}^s=(\pm \bm m^s_{r_A r_B}, \bm c^s_{r_A r_B}),
\end{equation}
where $\bm c_{r_Ar_B}^s$ is again given by Eq.~\eqref{eq:class}, with the values of $r_A$ and $r_B$ allowed by the constraint $\bm f_{ij}^s\cdot\bm c^s_{r_Ar_B}=2$, and $\bm m_{r_A r_B}^s=(1, (-1)^{r_A}, s, s(-1)^r_B)$. From the the fact that these facets are in one to one correspondence with the PR boxes, and that the latter cover the total number of nonlocal vertices of $\NC\SC$, follows inmmediatly that we have found all non trivial facets of the local polytope. Finally, it is worth noting that all the functionals $\bm\beta_{ij}^s$ are the same up to relabeling of inputs and outputs. We therefore come to the conclusion that, up to relabeling transformations, there is only one tight Bell inequality in the $(2,2)$ scenario, which is the celebrated CHSH inequality
\begin{equation}
    \mbraket{A_0B_0}+\mbraket{A_0B_1}+\mbraket{A_1B_0}-\mbraket{A_1B_1}\leq 2. \label{eq:CHSH}
\end{equation}

\section{Quantum nonlocal behaviors}

As discussed in Sec. \ref{sec:sets}, from the strict inclusion relation $\LC\subset\QC\subset\NC\SC$ between the correlation sets follows that there are probability points in $\QC$ that do not belong to $\LC$. Now that we have found all the non trivial facets of the polytope and realized that the CHSH inequality is the only tight Bell inequality in the (2, 2) scenario, we have a simple solution to the membership problem: a behavior $\bm q\in\QC$ admits a local realistic model iff inequality \eqref{eq:CHSH} is satisfied. 

A different, but closely related problem is that of maximizing the value $\beta$ of CHSH functional,
\begin{equation}
    \beta:= \mbraket{A_0B_0} + \mbraket{A_0B_1} + \mbraket{A_1B_0} - \mbraket{A_1B_1},
\end{equation}
over the set $\QC$ of quantum behaviors.

\section{Characteristic function}
The relabelling of the settings allows generalizing of the result from CHSH (2,2,2), namely two parties (Alice and Bob), two settings per party, and two outcomes to (n,2,2) scenario, as it has been proven in Ref\cite{WW01,ZB02}.
To generalize the relabelling process, we define the characteristic
function of the random settings, namely discrete Fourier transform
of the probability distribution:

\begin{equation}
\varphi\left(r_{1}r_{2}|xy\right)=\sum_{a,b\in\left\{ 0,1\right\} }\mathrm{e}^{\mathrm{i}\frac{2\pi}{d}\left(ar_{1}+br_{2}\right)}p\left(ab|xy\right),\qquad(d=2).
\end{equation}
Interesting for CHSH it is enough to consider $\varphi\left(11|xy\right)\equiv E^{xy}$,
so that
\begin{equation}\label{Exy}
E^{xy}=p\left(00|xy\right)-p\left(01|xy\right)-p\left(10|xy\right)+p\left(11|xy\right).
\end{equation}
Therefore for each setting we have
\begin{equation}
\mathbb{E}=\left(\begin{array}{cc}
E^{11} & E^{-11}\\
E^{1-1} & E^{-1-1}
\end{array}\right).
\end{equation}
Notice that in the case of deterministic assumption $E_{d}^{xy}=\left(-1\right)^{\ell}$,
for $\ell\in I$, because in Eq.~\eqref{Exy} only one term is different than zero. Up to a multiplicative constant, the entries of $\mathbb{E}$ are equivalent to the ones in $\bm c_d$ in Eq.~\eqref{joint_ind}. Therefore in a classical model $\mathbb{E}$ is given by
the convex combination ruled by $p_{\ell_1,\ell_2,\ell_3,\ell_4}\ge 0$,
\begin{equation}
\mathbb{E}=\sum_{i=1}^{4}\sum_{\ell_{i}\in I}p_{\ell_{1}\ell_{2}\ell_{3}\ell_{4}}\left(\begin{array}{c}
\left(-1\right)^{\ell_{1}}\\
\left(-1\right)^{\ell_{2}}
\end{array}\right)\otimes\left(\begin{array}{c}
\left(-1\right)^{\ell_{3}}\\
\left(-1\right)^{\ell_{4}}
\end{array}\right),\qquad\sum_{i=1}^{4}\sum_{\ell_{i}\in I}p_{\ell_{1}\ell_{2}\ell_{3}\ell_{4}}=1,\label{eq:EE}
\end{equation}
where the external product is defined as
\begin{equation}
\left(\begin{array}{c}
\alpha_{1}\\
\alpha_{2}
\end{array}\right)\otimes\left(\begin{array}{c}
\beta_{1}\\
\beta_{2}
\end{array}\right)=\left(\begin{array}{cc}
\alpha_{1}\beta_{1} & \alpha_{1}\beta_{2}\\
\alpha_{2}\beta_{1} & \alpha_{2}\beta_{2}
\end{array}\right).
\end{equation}
In this formalism, CHSH is obtained, up to a rescaling factor, by
choosing a basis orthogonal to the deterministic vectors in Eq.\ref{eq:EE}.
For instance,
\begin{equation}
q^{m_{1}m_{2}}=\left(\begin{array}{c}
1\\
\left(-1\right)^{m_{1}}
\end{array}\right)\otimes\left(\begin{array}{c}
1\\
\left(-1\right)^{m_{2}}
\end{array}\right).
\end{equation}
Indeed these bases $\mathcal{B}_1=\{(1,(-1)^{m_1})^T\}$ and $\mathcal{B}_2=\{(1,(-1)^{m_1})^T\}$ in this special case coincides and give always 2 or vanishing term in the modulus of the scalar product with a deterministic vector in Eq. \ref{eq:EE}:
\begin{equation}\label{gem}
    \left\langle
    \left(\begin{array}{c}
(-1)^{\ell_1}\\
(-1)^{\ell_2}
\end{array}\right),\left(\begin{array}{c}
1\\
(-1)^{m}\\
\end{array}\right)
\right\rangle=
\begin{cases}
2 & \ell_1=0, \ell_2+m=0\\
0 & \ell_1=0,1 \ell_2+m=1,0\\
-2 & \ell_1=1, \ell_2+m=1.
\end{cases}
\end{equation}
This yields
\begin{equation}\label{Eq4}
\left\langle \mathbb{E}|q\right\rangle _{\mathrm{HS}}\le4.
\end{equation}
This observation is the gem that brings all the family of complete Bell inequalities for the scenario $(n,2,2)$\cite{WW01,ZB02}. However here we do not normalize the two bases $\mathcal{B}_1$ and $\mathcal{B}_2$, because in a potential generalization one would reproduce the property in Eq. \eqref{gem}, by choosing a family of bases in $\mathbb{C}^d$, $\{\mathcal{B}^{(i)}\}_{i=1}^n$ with vectors $\{e_k^{(i)}\}_{k=1}^d$ which are mutually conjugated, i. e. $\langle e_k^{(i)}| e_\ell^{(j)}\rangle=\delta_{k\ell}$ for $i\neq j$. Trivially if the basis is normalized then the only mutual conjugated basis is itself.
This method allows to recover not only the Bell-CHSH, but also CGLMP\cite{Karczewski_2022}. As a byproduct, from a Bell inequality one can obtain a nonlocal witness. Suppose that a quantum experiment is prepared such that we can compute the probability from Eq. \eqref{qprob}
\begin{equation}
E^{xy}=\sum_{a,b\in \{0,1\}}\mathrm{e}^{\mathrm{i}\frac{2\pi}{d}(a+b)}p(ab|xy)=\sum_{a,b\in \{0,1\}}\mathrm{e}^{\mathrm{i}\frac{2\pi}{d}(a+b)}\mathrm{Tr}\left(M_x^a\otimes M_y^b \rho\right)=\mathrm{Tr}(B^{xy} \rho),
\end{equation}
where $B^{xy}$ is the Bell operator:
\begin{equation}
    B^{xy}=\sum_{a,b\in \{0,1\}}\mathrm{e}^{\mathrm{i}\frac{2\pi}{d}(a+b)}M_x^a\otimes M_y^b,\qquad d=2.
\end{equation}
Given that, the Eq.\eqref{Eq4} can be written as follows
\begin{equation}
    \sum_{x,y\in I}E^{xy}q^{xy}=\sum_{x,y\in I}\mathrm{Tr}(B^{xy}\rho)q^{xy}\leq 4 \mathrm{Tr}\rho\Rightarrow \mathrm{Tr}W\rho \ge 0
\end{equation}
with
\begin{align}
    W=&4 (\bm 1 \otimes \bm 1) - \sum_{x,y\in I}B^{xy}q^{xy}\nonumber\\
    =&4 (\bm 1 \otimes \bm 1) - \sum_{x,y\in I}\sum_{a,b\in \{0,1\}}\mathrm{e}^{\mathrm{i}\frac{2\pi}{d}(a+b)}M_x^a\otimes M_y^bq^{xy},
\end{align}
called \emph{non-local witness}. It is well-known\cite{Chruscinski2014} that, since the entanglement is a necessary condition for violation of Bell inequality, then a non-local witness is also an \emph{entanglement witness}\cite{Sarbicki_2020,Sarbickijpa_2020}.
\section{Discussion}
We revisited the fundamental aspect of the CHSH inequality, the most minimal Bell functionals that manifests a departure from the classical worldview. In particular, we show that up to a relabelling procedure it is possible to obtain all the facets of the \emph{Bell local} polytope. In particular, thanks to the symmetrical properties we restricted the geometrical properties to only a subpolytope shown in Fig. \ref{polytope} We present this promising aspect because it brought at the solution of (n,2,2) scenario, and this relieved some hopes for further generalizations. Therefore we mentioned possible avenues via the notion of conjugated basis. This is an interesting line of research for two reasons: given two conjugated bases, only one pair of the scalar product survive. This is a way to simplify the enormous numerical computation for more complicated scenarios, and the definition of conjugated basis is complementary to the definition of the well-studied and tricky mutually unbiased basis. Finally, we conclude that despite the qubit theory, involving the CHSH scenario, having been well studied for so long\cite{entropy_FID_CHSH}, it still constitutes a paradigm to assess the violation of the local realistic nature of bosonic systems of a different nature as the fields \cite{SP_1,SP_2,SP_PLA,CommentDun,SP_3,Non_contexAM}. Moreover, it reveals still nuances for contextual ontological models. Specifically, it is open the question of whether the $XZ$--\emph{uncertainty relation}\cite{Catani2022} can be recast into a CHSH-Bell scenario, similarly to parity oblivious random access code \cite{POM}.

\section*{Acknowledgments}
GS is supported by QuantERA/2/2020, an ERA-Net co-fund in Quantum Technologies, under the eDICT project. AM is supported 
by  Foundation for Polish Science (FNP), IRAP project ICTQT, contract no. 2018/MAB/5, co-financed by EU  Smart Growth Operational Programme, 
and by (Polish) National Science Center (NCN): MINIATURA  DEC-2020/04/X/ST2/01794.
NG is supported by the National Science
Centre, Poland, under the SONATA project “Fundamental aspects of the quantum set of correlations”, Grant
No. 2019/35/D/ST2/02014.

\bibliographystyle{ws-ijqi}

\begin{thebibliography}{99}

\bibitem{Bell1964}
J.~S. Bell, {\em Speakable and Unspeakable in Quantum Mechanics}, second
  edition edn. (Cambridge University Press, 1964).

\bibitem{Bell1964a}
J.~S. Bell, {\em Physics Physique Fizika} {\bf 1} (nov 1964) 195.

\bibitem{CHSH}
J.~F. Clauser, M.~A. Horne, A.~Shimony and R.~A. Holt, {\em Phys. Rev. Lett.}
  {\bf 23} (Oct 1969) 880.

\bibitem{CH74}
J.~F. Clauser and M.~A. Horne, {\em Phys. Rev. D} {\bf 10} (Jul 1974) 526.

\bibitem{Bavarian2015}
M.~Bavarian and P.~W. Shor, Information {Causality}, {Szemer\'edi}-{Trotter}
  and {Algebraic} {Variants} of {CHSH}, in {\em Proceedings of the 2015
  {Conference} on {Innovations} in {Theoretical} {Computer} {Science}\/},
  {ITCS} '15 (ACM, New York, NY, USA, 2015).

\bibitem{Peres2004}
A.~Peres and D.~R. Terno, {\em Reviews of Modern Physics} {\bf 76} (jan 2004)
  93.

\bibitem{Barrett}
J.~Barrett, L.~Hardy and A.~Kent, {\em Phys. Rev. Lett.} {\bf 95}  (2005) p.
  010503.

\bibitem{Cope2019}
T.~Cope and R.~Colbeck, {\em Physical Review A} {\bf 100} (aug 2019)

\bibitem{Pironio2014}
S.~Pironio, {\em Journal of Physics A: Mathematical and Theoretical} {\bf 47}
  (oct 2014) p. 424020.

\bibitem{Brunner2014}
N.~Brunner, D.~Cavalcanti, S.~Pironio, V.~Scarani and S.~Wehner, {\em Reviews
  of Modern Physics} {\bf 86} (apr 2014) 419.

\bibitem{Ziegler1995}
G.~M. Ziegler, {\em Lectures on Polytopes} (Springer New York, 1995).

\bibitem{Green1958}
J.~A. Green, {\em Mathematische Zeitschrift} {\bf 70} (dec 1958) 430.

\bibitem{Avis1992}
D.~Avis and K.~Fukuda, {\em Discrete \& Computational Geometry} {\bf 8} (sep
  1992) 295.

\bibitem{Weisstein2002}
E.~W. Weisstein, {\em CRC Concise Encyclopedia of Mathematics CD-ROM} (CRC,
  2002).

\bibitem{Khachiyan2008}
L.~Khachiyan, E.~Boros, K.~Borys, K.~Elbassioni and V.~Gurvich, {\em Discrete
  \& Computational Geometry} {\bf 39} (mar 2008) 174.

\bibitem{Fine1982}
A.~Fine, {\em Phys. Rev. Lett.} {\bf 48}  (1982) 291.

\bibitem{Pironio2005}
S.~Pironio, {\em Journal of mathematical physics} {\bf 46}  (2005) p. 062112.

\bibitem{Barrett1}
J.~Barrett, N.~Linden, S.~Massar, S.~Pironio, S.~Popescu and D.~Roberts, {\em
  Phys. Rev. A} {\bf 71}  (2005) p. 022101.

\bibitem{PR_orig}
S.~Popescu and D.~Rohrlich, {\em Foundations of Physics} {\bf 24}  (1994) 379.

\bibitem{scarani2019bell}
V.~Scarani, {\em Bell Nonlocality} (Oxford University Press, 2019).

\bibitem{Wood2015}
C.~J. Wood and R.~W. Spekkens, {\em New J. Phys.} {\bf 17} (mar 2015) p.
  033002.

\bibitem{chiribella2010probabilistic}
G.~Chiribella, G.~M. D'Ariano and P.~Perinotti, {\em Phys. Rev. A} {\bf 81}
  (June 2010) p. 062348.

\bibitem{acin2007device}
A.~Ac{\'\i}n, N.~Brunner, N.~Gisin, S.~Massar, S.~Pironio and V.~Scarani, {\em
  Physical Review Letters} {\bf 98}  (2007) p. 230501.

\bibitem{masanes2011secure}
L.~Masanes, S.~Pironio and A.~Ac{\'\i}n, {\em Nature communications} {\bf 2}
  (2011) 1.

\bibitem{kolmogorov1957elements}
A.~N. Kolmogorov and S.~V. Fomin, {\em Elements of the theory of functions and
  functional analysis. Volume 1: Metric and normed spaces} (Graylock Press
  Rochester, 1957).

\bibitem{Chaves2016}
R.~Chaves, {\em Phys. Rev. Lett.} {\bf 116} (Jan 2016) p. 010402.

\bibitem{Barret2005}
J.~Barrett, N.~Linden, S.~Massar, S.~Pironio, S.~Popescu and D.~Roberts, {\em
  Phys. Rev. A} {\bf 71} (Feb 2005) p. 022101.

\bibitem{WW01}
R.~F. Werner and M.~M. Wolf, {\em Phys. Rev. A} {\bf 64} (Aug 2001) p. 032112.

\bibitem{ZB02}
M.~\ifmmode~\dot{Z}\else \.{Z}\fi{}ukowski and C.~Brukner, {\em Phys. Rev.
  Lett.} {\bf 88} (May 2002) p. 210401.

\bibitem{Karczewski_2022}
M.~Karczewski, G.~Scala, A.~Mandarino, A.~B. Sainz and M.~Żukowski, {\em
  Journal of Physics A: Mathematical and Theoretical} {\bf 55} (sep 2022) p.
  384011.

\bibitem{Chruscinski2014}
D.~Chru\'sci\'nski and G.~Sarbicki, {\em Journal of Physics A: Mathematical and
  Theoretical} {\bf 47} (nov 2014) p. 483001.

\bibitem{Sarbicki_2020}
G.~Sarbicki, G.~Scala and D.~Chru{\'{s}}ci{\'{n}}ski, {\em Physical Review A}
  {\bf 101} (Jan 2020)

\bibitem{Sarbickijpa_2020}
G.~Sarbicki, G.~Scala and D.~Chru\'{s}ci\'{n}ski, {\em Journal of Physics A:
  Mathematical and Theoretical} {\bf 53} (oct 2020) p. 455302.

\bibitem{entropy_FID_CHSH}
A.~Mandarino and G.~Scala, {\em Entropy} {\bf 25}  (2023) p.~94.

\bibitem{SP_1}
T.~Das, M.~Karczewski, A.~Mandarino, M.~Markiewicz, B.~Woloncewicz and
  M.~{\.{Z}}ukowski, {\em New Journal of Physics} {\bf 24} (mar 2022) p.
  033017.

\bibitem{SP_2}
T.~Das, M.~Karczewski, A.~Mandarino, M.~Markiewicz, B.~Woloncewicz and
  M.~{\.{Z}}ukowski, {\em New Journal of Physics} {\bf 23} (jul 2021) p.
  073042.

\bibitem{SP_PLA}
T.~Das, M.~Karczewski, A.~Mandarino, M.~Markiewicz, B.~Woloncewicz and
  M.~Żukowski, {\em Physics Letters A} {\bf 435}  (2022) p. 128031.

\bibitem{CommentDun}
T.~Das, M.~Karczewski, A.~Mandarino, M.~Markiewicz, B.~Woloncewicz and
  M.~{\.{Z}}ukowski, {\em New Journal of Physics} {\bf 24} (mar 2022) p.
  038001.

\bibitem{SP_3}
T.~Das, M.~Karczewski, A.~Mandarino, M.~Markiewicz and M.~\ifmmode~\dot{Z}\else
  \.{Z}\fi{}ukowski, {\em Phys. Rev. Applied} {\bf 18} (Sep 2022) p. 034074.

\bibitem{Non_contexAM}
K.~Schlichtholz, A.~Mandarino and M.~{\.{Z}}ukowski, {\em New Journal of
  Physics} {\bf 24} (oct 2022) p. 103003.

\bibitem{Catani2022}
L.~Catani, M.~Leifer, G.~Scala, D.~Schmid and R.~W. Spekkens, {\em Physical
  Review Letters} {\bf 129} (dec 2022)

\bibitem{POM}
R.~W. Spekkens, D.~H. Buzacott, A.~J. Keehn, B.~Toner and G.~J. Pryde, {\em
  Phys. Rev. Lett.} {\bf 102}  (2009) p. 010401.

\end{thebibliography}

\end{document}